\documentclass[12pt]{article}
\pdfoutput=1
\usepackage{subfigure}
\usepackage{amssymb,amsmath}
\usepackage{color}
\usepackage{cancel}
\usepackage[colorlinks=true
,urlcolor=blue
,citecolor=blue
,linkcolor=blue
,pagecolor=blue
,linktocpage=true
,pdfproducer=medialab
]{hyperref}
\usepackage[a4paper,width=15.2cm]{geometry}

\usepackage{ulem}
\usepackage{placeins}

\usepackage{psfrag,epsf}
\usepackage{epsfig,eepic,epic,rotate}
\usepackage{graphicx,epsfig,color}
\usepackage{enumerate}
\usepackage{bm}
\usepackage{dcolumn}
\usepackage{url}
\usepackage{multirow}
\usepackage{rotating}
\usepackage{units}
\usepackage{epstopdf}
\usepackage{color}
\usepackage{pbox}
\newcommand{\be}{\begin{equation}}
\newcommand{\ee}{\end{equation}}

\newcommand{\bea}{\begin{eqnarray}}
\newcommand{\eea}{\end{eqnarray}}
\makeatletter \renewcommand{\@dotsep}{10000} \makeatother
\def\be{\begin{equation}}
\def\ee{\end{equation}}
\def\bea{\begin{eqnarray}}
\def\eea{\end{eqnarray}}
\def\bi{\begin{itemize}}
\def\ei{\end{itemize}}

\newcommand{\beq}{\begin{equation}}
\newcommand{\eeq}{\end{equation}}

\begin{document}
%Remove date before submitting to arXi\date{\today}

\begin{center}
{\Large\bf
Generalized Soft Breaking Leverage for the MSSM
}
{\vspace{0.3cm}

\Large
Cem Salih \"{Un}\footnote{
Email: cemsalihun@uludag.edu.tr},
\Large
\c{S}\"{u}kr\"{u} Hanif Tany\i ld\i z\i$^2$,
\Large
Saime Kerman$^3$,
\Large
Levent Solmaz$^4$
}

{\vspace{0.3cm}
{\it  $^{1}$\small{Department of Physics, Uluda\u{g} University, 16059, Bursa, T\"{u}rkiye}.}\\
{\it $^{2}$\small{Bogoliubov Laboratory of Theoretical Physics, Joint Institute for Nuclear Research, 141980, Dubna, Moscow Region, Russia}.} \\
{\it $^{3}$\small{Department of Physics, Dokuz  Eyl\"ul  University, 35210, \.Izmir, T\"{u}rkiye}.}\\
{\it $^{4}$\small{Department of Physics, Bal\i kesir University, 10145,
Bal\i kesir, T\"{u}rkiye}.}\\}

\section*{Abstract}
\end{center}
In this work we study  implications of  additional non-holomorphic soft breaking terms ($\mu^\prime$, $A^\prime_t$, $A^\prime_b$ and $A^\prime_\tau$) on the MSSM phenomenology. By respecting the existing bounds on the mass measurements and restrictions coming from certain B-decays, we probe reactions of the MSSM to these additional soft breaking terms. We provide examples in which some slightly excluded
solutions of the MSSM can be made to be consistent with the current experimental results. During this, even after applying additional fine-tuning constraints the non-holomorphic terms are allowed to be as large as hundreds of GeV. Such terms prove that they are capable of enriching the phenomenology and varying the mass spectra of the MSSM heavily, with a reasonable amount of fine-tuning.

We observe that higgsinos, the lightest stop,
the heavy Higgs boson states $A,H,H^{\pm}$, sbottom and stau exhibit
the highest sensitivity to the new terms. We also show how the light
stop can become  nearly degenerate with top quark using these non-holomorphic terms.

\newpage
\section{Introduction}
\label{sec:intro}

Despite the excitement of the Higgs boson discovery in ATLAS \cite{ATLAS} and
 CMS \cite{CMS}, the results from the experiments conducted at the Large Hadron Collider (LHC)
 have brought a severe pressure on the supersymmetric models. Indeed,
  there have been no signal from the supersymmetric partners of the Standard Model (SM) particles.
   While motivations for SUSY did not disappear, a 125 GeV  SM-like Higgs boson requires rather heavy stops that
    leads to the fine-tuning problem in the minimal supersymmetric extension of the SM (MSSM).
     Additionally, the LHCb results for the rare decays of B meson have a significant impact on the parameter
      space of the supersymmetric models such as constrained MSSM (CMSSM) and non-universal Higgs mass models (NUHM) \cite{Roszkowski:2014wqa}.
      For instance the observation of $B_{s}\rightarrow \mu^{+}\mu^{-}$ \cite{Aaij:2012nna} and
      the updated range of $B\rightarrow X_{s}\gamma$ \cite{Amhis:2012bh} especially disfavor CMSSM.

The scrutiny within the supersymmetric models may consider the lack of evidences
 to be incompleteness of such models, since supersymmetry (SUSY) has strong motivations
 such as resolution of the gauge hierarchy problem \cite{Barbieri:1987fn}, unification of the gauge couplings \cite{Georgi:1974sy},
 radiative electroweak symmetry breaking (REWSB) \cite{Higgs:1964pj}, dark matter candidate under R-parity conservation, etc.
  Considering the strong impacts of the experimental results, extensions of the MSSM such as next to MSSM (NMSSM) \cite{Fayet:1974pd},
   R-parity violation (RPV) \cite{Hall:1983id} have been excessively investigated and it has been found that
    such extended models are capable of providing results at the low energy scale that are in much better fit to the experimental results.

Alternatively and arguably as a much simpler way to extend the MSSM, one also can examine the generalized MSSM by considering non-holomorphic (NH) terms in the soft supersymmetry breaking (SSB) sector of the theory \cite{Jack:1999ud}. For simplicity, we restrict our search to the MSSM domain, but the consideration can be enlarged to the extended models \cite{Demir:2005ti}. In addition to the MSSM soft breaking terms, the following terms exist in the NH extension of MSSM (NHSSM).

\begin{eqnarray}
\label{nonholSSB}
{\cal{L}}_{soft}^{\prime} =\mu^\prime {\tilde H_u}\cdot
{\tilde H_d} +\tilde{Q}~{H}_d^{\dagger} A^\prime_{u} \tilde{U}+
\tilde{Q}~ {H}_u^{\dagger} A^\prime_{d} \tilde{D}
 + \tilde{L}~{H}_u^{\dagger} A^\prime_{e} \tilde{E} + \mbox{h.c.}
\end{eqnarray}
where $\mu'$ is the Higgsino mixing term, and $A'_{u,d,e}$ are NH trilinear scalar couplings. We use a similar notation to the holomorphic supersymmetric Lagrangian, but $\mu'$ and $A'_{u,d,e}$ are independent of the holomorphic terms and treated as the free parameters of NHSSM. This similar notation is based on the fact that we do not add any new particle to the MSSM content,
  but rather we assume only the existence of NH terms given above. During our numerical investigation,
  we also assume  CP and the R-parity to be conserved and require our solutions to satisfy that the lightest supersymmetric particle (LSP) is the lightest neutralino.

As can be predicted, the additional terms given in Eq.(\ref{nonholSSB}) can yield in quite different phenomenology at the low scale. Since the degree of freedom is greater than the MSSM, the region of the parameter space consistent with the current experimental constraints can be found much larger in NHSSM than that found in MSSM. To see this, let us start with the NH contributions to the supersymmetric mass spectrum, which can be summarized for scalar fermions as follows \cite{Jack:1999ud}:

\begin{equation}
 M_{\tilde{f}}^{2} = \left( \begin{array}{cc}
 m_{\tilde{f}_{L}\tilde{f}^{*}_{L}} & X_{\tilde{f}} \\ \\
X_{\tilde{f}}^{*} & m_{\tilde{f}_{R}\tilde{f}^{*}_{R}}
 \end{array} \right)
 \label{sfermionsmass2}
 \end{equation}
Here $M^{2}_{\tilde{f}}$ is the general form of the mass-squared mass matrices of sfermions written in basis ($\tilde{f}_{L}, \tilde{f}_{R}$) and ($\tilde{f}^{*}_{L}, \tilde{f}^{*}_{R}$) where $\tilde{f} = \tilde{u}, \tilde{d},\tilde{e}$ stands for up-type squarks, down-type squarks and sleptons respectively. The masses and mixings of sfermions can be written as follows:

\begin{eqnarray}
\label{defs}
m_{\tilde{u}_{L}\tilde{u}^{*}_{L}} = -\frac{1}{24}(-3g_{2}^{2}+g_{1}^{2})(-v_{u}^{2}+v_{d}^{2})+\frac{1}{2}(2m_{q}^{2}+v_{u}^{2}Y_{u}^{\dagger}Y_{u}),  \nonumber \\
m_{\tilde{u}_{R}\tilde{u}^{*}_{R}}= \frac{1}{2}(2m_{u}^{2}+v_{u}^{2}Y_{u}Y_{u}^{\dagger})+\frac{1}{6}g_{1}^{2}(-v_{u}^{2}+v_{d}^{2}),~~~~~~~~~~~~~~~~\nonumber \\
X_{\tilde{u}} = -\frac{1}{\sqrt{2}}[v_{d}(\mu Y^{\dagger}_{u}+A^{'\dagger}_{u})-v_{u}A_{u}^{\dagger}],~~~~~~~~~~~~~~~~~~~~~~~~~~~~ \nonumber \\
\hline \nonumber \\
 m_{\tilde{d}_{L}\tilde{d}^{*}_{L}} = -\frac{1}{24}(3g_{2}^{2}+g_{1}^{2})(-v_{u}^{2}+v_{d}^{2})+\frac{1}{2}(2m_{q}^{2}+v_{d}^{2}Y_{d}^{\dagger}Y_{d}),~~~~ \nonumber \\
 m_{\tilde{d}_{R}\tilde{d}^{*}_{R}}= \frac{1}{2}(2m_{d}^{2}+v_{d}^{2}Y_{d}Y_{d}^{\dagger})+\frac{1}{12}g_{1}^{2}(-v_{d}^{2}+v_{u}^{2}),~~~~~~~~~~~~~~~~  \\
X_{\tilde{d}} = -\frac{1}{\sqrt{2}}[v_{u}(\mu Y_{d}^{\dagger}+A^{'\dagger}_{d})-v_{d}A_{d}^{\dagger}],~~~~~~~~~~~~~~~~~~~~~~~~~~~~~~~~ \nonumber \\
\hline \nonumber \\
m_{\tilde{e}_{L}\tilde{e}^{*}_{L}} = \frac{1}{2}v_{d}^{2}Y^{\dagger}_{e}Y_{e}+\frac{1}{8}(-g_{2}^{2}+g_{1}^{2})(-v_{u}^{2}+v_{d}^{2})+m_{l}^{2},~~~~~~~~~~~ \nonumber \\
m_{\tilde{e}_{R}\tilde{e}^{*}_{R}} =\frac{1}{2}v_{d}^{2}Y_{e}Y_{e}^{\dagger}+\frac{1}{4}g_{1}^{2}(-v_{d}^{2}+v_{u}^{2})+m_{e}^{2},~~~~~~~~~~~~~~~~~~~~~~ \nonumber \\
X_{\tilde{e}} = \frac{1}{\sqrt{2}}[-v_{u}(\mu Y_{e}^{\dagger}+A^{'\dagger}_{e})+v_{d}A_{e}^{\dagger}].~~~~~~~~~~~~~~~~~~~~~~~~~~~~~~~~~ \nonumber \\
\hline \nonumber
 \end{eqnarray}
 
\begin{figure}[htp!]
\centering
\includegraphics[scale=0.30]{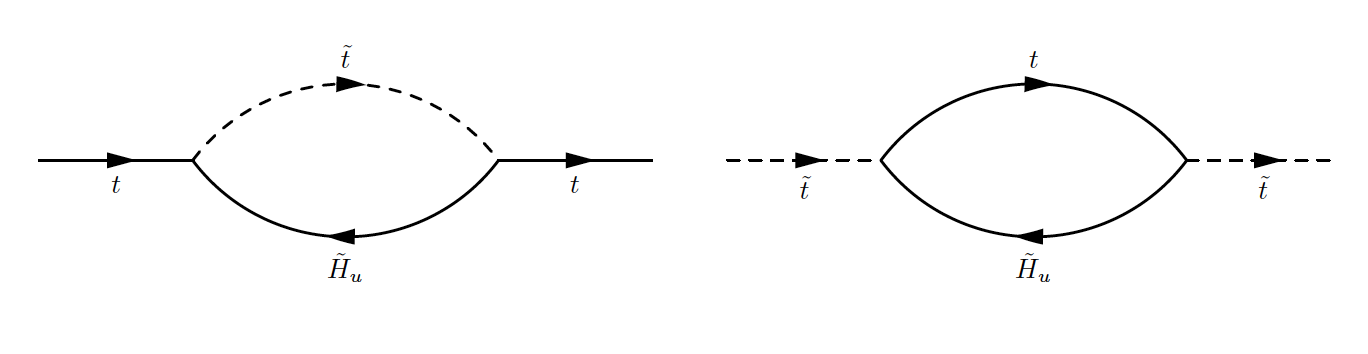}
\caption{Self-energy diagrams for top quark and squarks involving with the Higgsinos.}
\label{nonholdiags}
\end{figure} 
 
Even though the diagonal elements are well-known masses of sfermions, the NH terms appear in the off-diagonal elements and hence they can significantly change the sfermion masses by altering their mixings. Note that $\mu'-$term does not appear in the scalar masses at tree level, since it is introduced to the Lagrangian only with Higgsinos. The MSSM Lagrangian introduces the Yukawa interactions between the fermions and sfermions through the Higgsino vertices, in addition to the Higgs-fermion-fermion Yukawa interactions of the Standard Model \cite{Martin:1997ns}, and such vertices contribute to masses of quarks and squarks at the loop level \cite{Hollik:2014wea}. Figure \ref{nonholdiags} shows some of such diagrams for the top quark and squark. We suppress the handedness subscripts, but the diagrams are drawn with the necessary conservations including the R-parity. The similar diagrams can be repeated for the other quarks and leptons. In the MSSM framework in which the non-holomorphic terms do not exist, such contributions are controlled by the well-known holomorphic $\mu$ and $A_{t,b,\tau}$ terms. On the other hand, in NHSSM, such non-holomorphic terms are not set to zero at tree level, and the higgsinos are also controlled by the non-holomorphic $\mu'-$term. Hence it contributes to masses of sfermions at loop level. Considering the large Yukawa couplings associated with the third family, one can expect significant effects on the third family sfermion masses from $\mu'-$term, even though there is no contribution from $\mu'$ at tree level. Similar discussion can be followed for the Higgs sector of the MSSM. Indeed, the non-holomorphic effects are not seen directly, since the tree level Higgs potential is the same as that in the MSSM framework. In a conventional approach, one can derive the tree level Higgs masses only with two parameters, say the mass of CP-odd Higgs $m_{A}$ and $\tan\beta$, in the Higgs sector of MSSM. On the other hand, considering the higher order diagrams involving with the Higgses, for instance self energy diagrams \cite{Hollik:2014wea,Frank:2013hba}, $\mu'-$term contributes to the masses of the Higgs bosons through higgsino loops. In addition to $\mu'-$term, the NH trilinear scalar interaction terms, $A^\prime_{t,b,\tau}$ contribute to the Higgs masses at loop levels \cite{Cincioglu:2009qm}. Such contributions can have important results for the fine-tuning \cite{Kitano:2005wc}, since the 125 GeV Higgs boson mass can be satisfied without having heavy stops or large mixing in contrast to the case of MSSM \cite{Carena:2011aa}.

Similarly the square mass matrices for the neutralino and chargino can be written as:

\begin{equation}
M_{\tilde{\chi}^{0}} = \left(  \begin{array}{cccc}
M_{1} & 0 & -\frac{1}{2}g_{1}v_{d} & \frac{1}{2}g_{1}v_{u} \\
0 & M_{2} & \frac{1}{2}g_{2}v_{d} & -\frac{1}{2}g_{2}v_{u} \\
-\frac{1}{2}g_{1}v_{d} & \frac{1}{2}g_{2}v_{d} & 0 & -\mu+\mu^\prime \\
\frac{1}{2}g_{1}v_{u} & -\frac{1}{2}g_{2}v_{u} & -\mu+\mu^\prime & 0
\end{array} \right)
\label{neutcharmass2}
\end{equation}
and
\begin{equation}
M_{\tilde{\chi}^{\pm}} = \left(  \begin{array}{cc}
M_{2} & \frac{1}{\sqrt{2}}g_{2}v_{u} \\
\frac{1}{\sqrt{2}}g_{2}v_{d} & -\mu^\prime+\mu
\end{array}  \right),
\label{charmass2}
\end{equation}
where $M_{\tilde{\chi}^{0}}$ is mass matrix for the neutralinos in
the basis ($\tilde{B},\tilde{W}^{0},
\tilde{H}_{d}^{0},\tilde{H}_{u}^{0}$) and ($\tilde{B},\tilde{W}^{0},
\tilde{H}_{d}^{0},\tilde{H}_{u}^{0}$), while
$M_{\tilde{\chi}^{\pm}}$ is for the charginos in the basis
$(\tilde{W}^{-},\tilde{H}_{d}^{-})$ and
$(\tilde{W}^{+},\tilde{H}_{u}^{+})$. While all the NH terms affect
sfermion masses, only $\mu'-$term is effective in the neutralino and
chargino sector at tree-level. It is easy to infer from
Eqs.(\ref{neutcharmass2},\ref{charmass2}) that the lightest mass
eigenvalues of neutralino and chargino mass matrices are to be very
small when $\mu^\prime \approx \mu$. In this
context, the NH terms can yield almost massless higgsino-like LSP.

In this paper, we explore the low scale phenomenology in the NHSSM framework, and we consider effects of the NH terms by considering two benchmark points. We aim to probe allowed parameter space of the NHSSM in accord with the current experimental constraints. The outline of the rest of the paper is as follows. We explain the scanning procedure and the experimental constraints applied in our analysis in Section \ref{sec:scan}, where we also briefly describe the benchmark points and their implications in MSSM. We present the results and phenomenological determination of ranges of the NH terms in Section \ref{sec:pheno}. We devote Section \ref{sec:finetuning} on a few words on the fine-tuning in NHSSM, and finally; we summarize and conclude our results in Section \ref{sec:conc}.

\section{Scanning Procedure}
\label{sec:scan}

In our approach, we focus on the low scale implications of the generalized MSSM in which the Lagrangian includes also the non-holomorphic terms mentioned in the previous section. As is well known, the MSSM has more than a hundred free parameters at the low scale. Instead of random determination, we set these free parameters respectively to the low scale predictions of two benchmark points which are obtained in the CMSSM framework as listed in Table 1 with their CMSSM input parameters. These points provide solutions for which the lightest neutralino is LSP, and the radiative electroweak symmetry breaking (REWSB) is satisfied. We employ state of the art codes which are the fortran code prepared by SARAH \cite{Staub:2013tta} for the use of SPheno \cite{Porod:2003um}. Also we set $\mu > 0$ and $m_{t}=173.3$ GeV \cite{ATLAS:2014wva}, where $m_{t}$ is the mass of top quark. Note that one or two sigma variation in $m_{t}$ do not change the results too much \cite{Gogoladze:2011db}. Once we recalculate the low scale observables in our scan after taking into account the contributions from the NH terms, we require  our solutions to satisfy the mass bounds \cite{Nakamura:2010zzi}, the constraints from the rare decays $B_{s}\rightarrow \mu^{+}\mu^{-}$ \cite{Aaij:2012nna} and $B\rightarrow X_{s}\gamma$ \cite{Amhis:2012bh}. These constraints can be summarized as follows:

\begin{eqnarray}
 \label{constraints}
m_{h}=(123-127)~{\rm GeV} \nonumber \\
m_{\tilde{g}} \geq 1.4~{\rm TeV} \nonumber \\
0.8\times10^{-9} \leq {\rm BR}(B_s \rightarrow \mu^+ \mu^-) \leq 6.2\times 10^{-9}~(2\sigma)  \\
2.99\times 10^{-4} \leq {\rm BR}(B\rightarrow X_{s}\gamma) \leq 3.87\times 10^{-4}~(2\sigma) \nonumber
\end{eqnarray}
where we display the current mass bounds on the SM-like Higgs boson \cite{ATLAS,CMS,Asner:2010qj} and gluino \cite{Aad:2012fqa}, because they have changed since the LEP era. We do not apply the Higgs mass bound strictly by taking it about 125 GeV, since the theoretical uncertainties in minimization of the scalar potential and the experimental uncertainties in measures of $m_{t}$ and $\alpha_{s}$ lead to about 3 GeV uncertainty in estimation of the Higgs boson mass. Note that the Higgs boson mass constraint has a strong impact on the stop sector, since it requires either heavy stops or large SSB trilinear $A_{t}-$term that lead to the stop masses at the order of TeV \cite{Djouadi:2005gj}. In addition to the constraints given in Eq.(\ref{constraints}), we require our solutions to do no worse than the SM prediction for the muon anomalous magnetic moment $\Delta(g-2)_\mu>0$, and we also imposed chargino LEP bound $m_{\chi^\pm}>105$ GeV.

\begin{table}[t!] \hspace{-2.0cm}
\centering
\scalebox{0.85}{
\begin{tabular}{|c|cc|}
\hline
\hline
&&\\
             MSSM    & BMP1 & BMP2  \\
               &&\\
\hline
$m_{0}$       & 749.6  & 1700    \\
$M_{1/2} $    & 986.2 & 425   \\
$\tan\beta$   & 29.7 & 15   \\
$A_0$         & -2450 & -3500    \\
$m_t$         & 173.3 & 173.3   \\
\hline
$A_{t}$ & -2082 & -1672\\
$A_{b}$ & -1439 & -807.2\\
$A_{\tau}$ & -771.2 & -539\\
$\mu$          & 1658  & 1478   \\
\hline
$m_h$            & 125.2 & 124.3    \\
$m_H$           & 1512  & 2038  \\
$m_A$           & 1506 & 2029   \\
$m_{H^{\pm}}$   & 1515 & 2039  \\

\hline
$m_{\tilde{\chi}^0_{1,2}}$
                 & \textbf{425}, 807.7  & \textbf{182.8}, 356.8 \\

$m_{\tilde{\chi}^0_{3,4}}$
                 & 1653, 1656 & 1477, 1480  \\

$m_{\tilde{\chi}^{\pm}_{1,2}}$
                & 807.9, 1656.8 & \textbf{357}, 1480   \\

$m_{\tilde{g}}$  & 2189 & 1088   \\
\hline $m_{ \tilde{u}_{1,2}}$
                 & 2104, 2104  & 1894, 1894   \\
$m_{\tilde{t}_{1,2}}$
                 & 1294, 1753 & \textbf{490.4}, 1379  \\
\hline $m_{ \tilde{d}_{L,R}}$
                 & 2105, 2105 & 1895, 1895   \\
$m_{\tilde{b}_{1,2}}$
                 & 1710, 1880 & 1349, 1810   \\
\hline
$m_{\tilde{\nu}_{e,\mu}}$
                 &1004, 1004 & 1718, 1718  \\
$m_{\tilde{\nu}_{\tau}}$
                 & 901.1 &  1679   \\
\hline
$m_{ \tilde{e}_{1,2}}$
                & 804, 913 & 1702, 1702   \\
$m_{ \tilde{\mu}_{1,2}}$
                & 1008, 1008 & 1720, 1720   \\
$m_{\tilde{\tau}_{1,2}}$
                & \textbf{490.1}, 803.3 & 1619, 1684   \\
\hline
${\rm BR}(B_{s}\rightarrow \mu^{+}\mu^{-})$ & $3.89\times 10^{-9}$ & $3.50\times 10^{-9}$   \\
${\rm BR}(B\rightarrow X_{s}\gamma)$ & $2.89\times 10^{-4}$ & $2.81\times 10^{-4}$   \\
$\Delta_{EW}$ &661.5 & 525.4 \\
\hline
\hline
\end{tabular}}
\caption{Benchmark points excluded by the constraints from the decay process $B\rightarrow X_{s}\gamma$ in the MSSM. All masses are given in GeV. The first block at top represents the GUT scale parameters, while all other blocks list the parameters at the low scale. Point 1 displays a solution with stau NLSP, while it is the lightest chargino in Point 2. Point 2 also depicts a solution with the lightest stop of mass about 490 GeV. The fine-tuning measures are in acceptable range ($\Delta_{EW} \lesssim 10^{3}$) for both points.}
\label{MSSMbenchmarks}

\end{table}

We present our benchmark points  in Table
\ref{MSSMbenchmarks} where all masses are given in GeV. Both points
satisfy REWSB and neutralino being LSP condition and they have
acceptable fine-tuning ($\Delta_{EW} \lesssim 10^{3}$) in the MSSM
framework. Point 1 is taken from Ref. \cite{Burgess:2012tq} and it
is currently excluded by the constraint from the rare decay process
$B\rightarrow X_{s}\gamma$. Point 1 is taken
as a sample to show contributions from the NH
Lagrangian of Eq.(\ref{nonholSSB}) and explain the cuts which we
apply to determine the ranges of the NH terms. In
addition to Point 1, we consider also Point 2 that is obtained from
our scan searching for light stops of mass about 500 GeV. It is
excluded by the  ${\rm BR}(B\rightarrow X_{s}\gamma)$ constraint like
Point 1. It also leads to the stop quark of 490 GeV mass that is
almost excluded for the LSP of mass about 180 GeV
\cite{Outschoorn:2013pma}. We aim for this point to lower the stop
mass with contributions from the NH terms down to
$\lesssim 200$ GeV whereby it is nearly degenerate with the top quark.

The motivation for the stop mass nearly degenerate with the top quark comes from the fact that the LHC has not excluded such light stop solutions yet \cite{Outschoorn:2013pma} and the recent studies \cite{Buckley:2014fqa} show that $\tilde{t}\tilde{t}^{*}$ cross section is less than the error in calculation of top pair production which is measured to be \cite{Collaboration:2013bma}

\begin{equation}
\sigma_{tt^{*}}^{\sqrt{s}=8~{\rm TeV}} = 241\pm 2~({\rm stat.}) \pm 31~({\rm syst.})\pm 9~({\rm lumi.}) ~{\rm pb}.
\label{toppair}
\end{equation}

When stop is almost degenerate with the top quark, decay products
from $t\bar{t}$ and $\tilde{t}\tilde{t}^{*}$ are identical and it is
challenging to distinguish stop and top quarks from each other
\cite{Buckley:2014fqa}. It has been also shown that it is possible
to obtain light stop masses about $\lesssim 200$ GeV in CMSSM,
however; a huge amount of fine-tuning is required due to a large
mixing between stop quarks in order to induce a 125 GeV Higgs boson
mass \cite{Demir:2014jqa}. There are exclusive studies which show
that requiring acceptable fine-tuning measures bound the stop mass
to about 500 GeV from below \cite{Hall:2011aa}. It is worth to study
with Point 2 in the NHSSM framework, because contributions from the
NH terms help to raise the Higgs boson mass and loose
stress on the stop sector. We explore the NH parameter
space in which the stop can be found to be nearly degenerate with
top quark and consistent with the fine-tuning constraints.

\section{Phenomenological Cut-offs for NH Contributions}
\label{sec:pheno}
We divided this section into  pieces in order to  emphasize the effects of NH terms separately.
We start with probing the impact of the $\mu^\prime$ term first by setting $A^\prime_{t,b,\tau}=0$. Then, the following subsection studies NH trilinear scalar interaction couplings.
\subsection{$\mu^\prime$ term}
\label{muprime}
Let us start to investigate  contributions from the non-holomorphic terms and phenomenological bounds on them by considering Point 1 of Table \ref{MSSMbenchmarks}, which is already inconsistent with the constraints from the rare decays of B-meson at $2\sigma$. Since the contributions from stop-chargino and the MSSM Higgs sector count for the supersymmetric contributions to such rare decays, one can expect that the NH mixing term, $\mu'$, can significantly change the B-physics implications.

\begin{figure}[h!]
\begin{center}
\includegraphics[scale=1,height=3.1cm,angle=0]{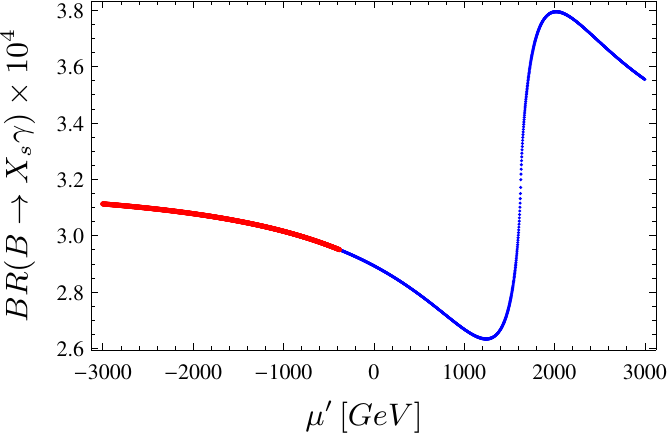}
\hfill
\includegraphics[scale=1,height=3.1cm,angle=0]{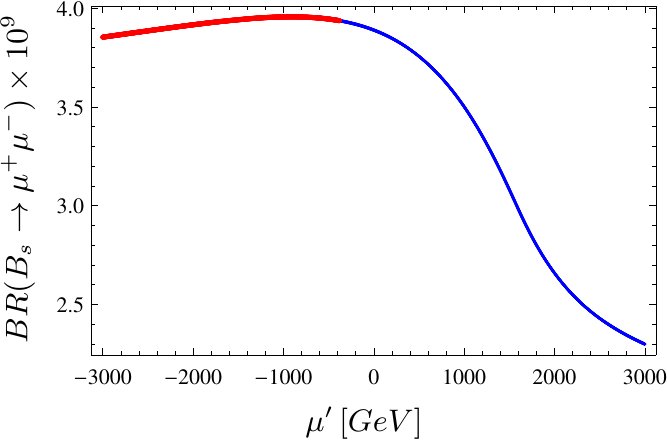}
\hfill
\includegraphics[scale=1,height=3.1cm,angle=0]{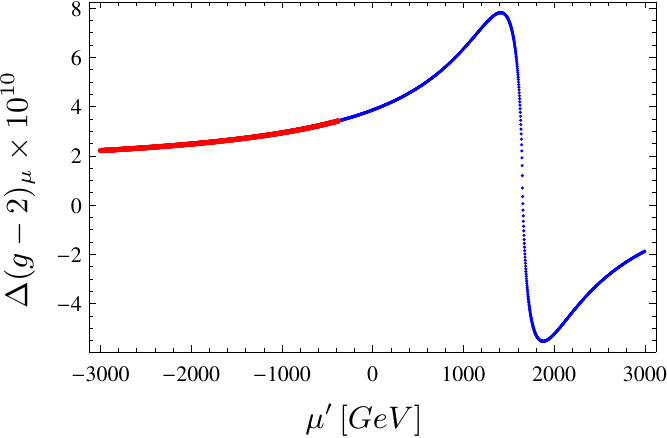}
\end{center}
\vspace{-0.6cm} \caption{Plots in $BR(B\rightarrow
X_{s}\gamma)-\mu^\prime$, $BR(B_{s}\rightarrow
\mu^{+}\mu^{-})-\mu^\prime$ and $\Delta(g-2)_\mu-\mu^\prime$ panels.
The plots are obtained for $A'_{t,b,\tau}=0$. The red part of the
curve represent the solutions which are consistent with the
experimental constraints mentioned in Section \ref{sec:scan}, while
the blue part is excluded. It should be noticed that
$A^\prime_{t,b,\tau}=0$ and $\mu^\prime=0$  corresponds to our BMP1.}
\label{figure1}
\end{figure}

Figure \ref{figure1} displays the plots in $BR(B\rightarrow
X_{s}\gamma)$, $BR(B_{s}\rightarrow \mu^{+}\mu^{-}) $ and
$\Delta(g-2)_\mu$ versus $\mu^\prime$  panels respectively. The
plots are obtained for $A'_{t,b,\tau}=0$. The red part of the curve
represent the solutions which are consistent with the experimental
constraints mentioned in Section \ref{sec:scan}, while the blue part
stands for being excluded. The NH contribution to the process
$B\rightarrow X_{s}\gamma$ can be written as $BR(B\rightarrow
X_{s}\gamma) \propto A_{t} - (\mu - \mu' + A'_{t})\cot\beta$
\cite{Cincioglu:2009qm} and for $A'_{f}=0$ we see that $\mu'
\lesssim -400$ GeV can provide enough contribution to satisfy the
constraint from $BR(B\rightarrow X_{s}\gamma)$ decay. The least
$BR(B\rightarrow X_{s}\gamma)$ prediction is obtained when $\mu' \approx 1.3$
TeV which happens in the blue region excluded by also several
constraints. On the other hand, one can obtain enough contribution
to $BR(B\rightarrow X_{s}\gamma)$ when $\mu' \approx 1600$ GeV,
however; it is excluded mostly by the bounds on the sparticle
masses. From  the middle panel of Figure \ref{figure1} one can read
how softly $BR(B_{s}\rightarrow \mu^{+}\mu^{-})$ prediction varies
with $\mu^\prime$ parameter. We see that this restriction is not
as strong as the one from $(B\rightarrow X_{s}\gamma)$ to bound
the related NH term for our BMP1. The last panel of Figure
\ref{figure1} represents the contributions to the muon anomalous
magnetic moment (muon $g-2$). The red region shows a slight decrease
in the $\Delta(g-2)_{\mu}$ while it remains in the acceptable range.

\begin{figure}[h!]
\includegraphics[scale=1,height=4.6cm,angle=0]{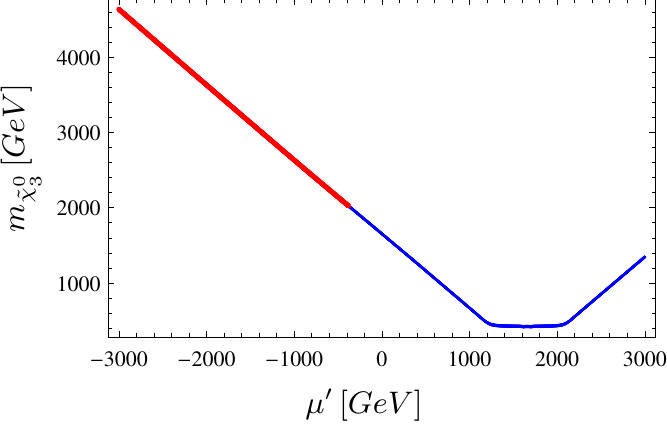}
\hfill
\includegraphics[scale=1,height=4.6cm,angle=0]{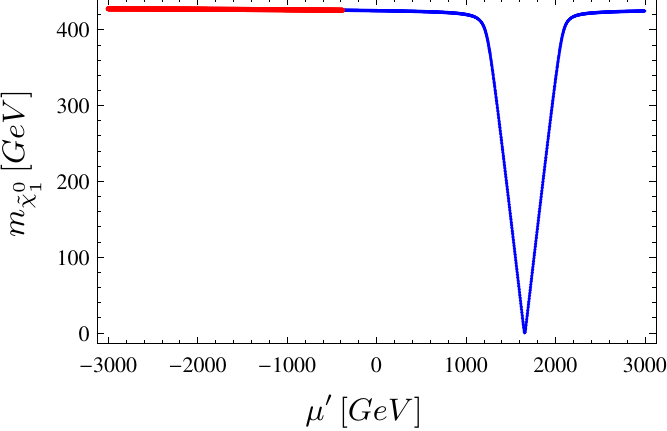}
\caption{Plots in $m_{\tilde{\chi}_{3}^{0}}-\mu'$ and $m_{\tilde{\chi}_{1}^{0}}-\mu'$ planes. The color coding is the same as Figure \ref{figure1}.}
\label{figure2}
\end{figure}

\begin{figure}[h!]
\begin{center}
\includegraphics[scale=1,height=4.6cm,angle=0]{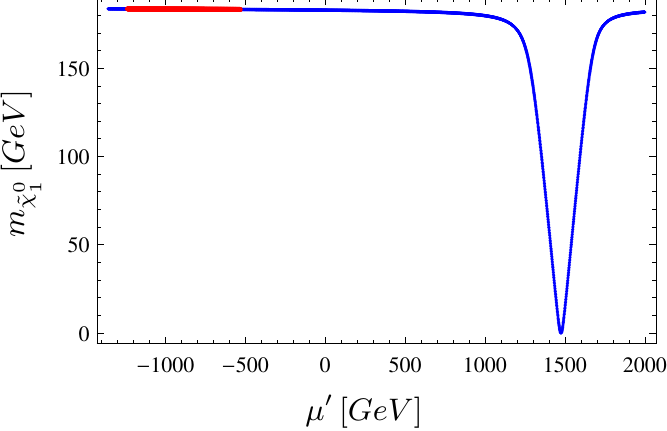}
\hfill
\includegraphics[scale=1,height=4.6cm,angle=0]{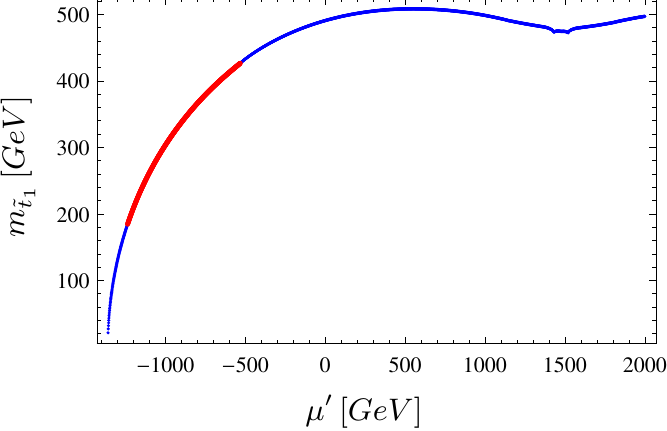}
\end{center}
\vspace{-0.4cm} \caption{Lightest neutralino and light stop masses
against $\mu^\prime$ for BMP2. The color coding is the same as
Figure \ref{figure1}.} \label{figure3}
\end{figure}

The NH contribution to the $BR(B\rightarrow X_{s}\gamma)$ and $\Delta(g-2)_{\mu}$ can be understood clearer if one considers the masses
 of neutralinos and charginos. As is mentioned above, when $\mu' \approx \mu = 1658$ for BMP1, the lightest neutralino mass tends to be
 zero as seen from plots in  $m_{\tilde{\chi}_{3}^{0}}-\mu'$ and $m_{\tilde{\chi}_{1}^{0}}-\mu'$ planes of Figure \ref{figure2}.
 The color coding is the same as Figure \ref{figure1}. In the CMSSM framework, the lightest neutralino is usually mostly bino, and the Higgsino components of neutralino
 are found to be relatively heavier. The $m_{\tilde{\chi}_{3}^{0}}-\mu'$ plane of Figure \ref{figure2} shows that the Higgsino mass linearly increases as $\mu'$ increases
  in the red region. However in the blue region with $1200 \lesssim \mu' \lesssim 2000$ GeV, $m_{\tilde{\chi}_{3}^{0}}$ remains constant even if $\mu'$ changes.
  It should be remembered that $\mu^\prime$ can drive masses of the lightest neutralino and chargino to zero when $\mu' \approx \mu=1658$ GeV for BMP1.
  We present lightest neutralino mass variation in the $m_{\tilde{\chi}_{1}^{0}}-\mu'$ plane (right panel) of  Figure \ref{figure2} for BMP1.
  One can easily see that the $\mu'-$term has no effect on the lightest neutralino mass in red region at all.
   It is because, the lightest neutralino is mostly bino in this region. However, when $\mu' \approx \mu$, the higgsinos become lighter than the bino and the lightest neutralino
   is formed mostly by the higssinos. A similar mass pattern is obtained also for the chargino sector. While the lightest chargino is mostly wino in CMSSM,
   it is found to be mostly higgsino in our model when $\mu' \approx \mu$. In this context, since the lightest chargino mass is close to zero, it is excluded by
   the LEP bound on chargino mass that is why it is observed in the blue part of the curves.
    As is also seen from $BR(B\rightarrow X_{s}\gamma)-\mu'$,  and $\Delta(g-2)_\mu - \mu'$ planes of Figure \ref{figure1},
    we obtain the steepest part of the curves in the same region with $\mu' \approx \mu$. Since it is very light, the chargino channel dominates over the
     supersymmetric contribution to $BR(B\rightarrow X_{s}\gamma)$ in this region. Similarly $\Delta(g-2)_{\mu}$ receives the
     dominant contributions from the neutralino-smuon channel. Note that the sign of contributions
      to $\Delta(g-2)_{\mu}$ is proportional to ${\rm sgn}((\mu-\mu')\times M_{2})$, and since $(\mu - \mu')$ changes
       its sign from positive to negative, the implications for $\Delta(g-2)_{\mu}$ become worse than the SM and hence it is
       excluded by our requirement that we assume the solutions to do no worse than the SM on $\Delta(g-2)_{\mu}$.
       In this context our requirement can bound the NH $\mu'-$term range in a general scan as $\mu' \lesssim \mu$. The situation is very similar as can be seen from the first panel of Figure \ref{figure3} for our BMP2.

As bounding the $\mu'-$term from above, one can also bound it from
below. As a comparison we present the lightest neutralino and
light stop masses against $\mu^\prime$ for BMP2 in Figure
\ref{figure3}. The color coding is the same as Figure \ref{figure1}.
A similar curve for the lightest neutralino mass is obtained when
$\mu' \approx \mu = 1478$ GeV. As shown in $m_{\tilde{t}_{1}}-\mu'$
plane, $\mu'$ leads to relatively lighter stop masses, and while the
stop mass is about 500 GeV in the CMSSM framework, it can be as
light as $\sim 180$ GeV in NHSSM. However, the blue curve takes over
the red one when $\mu' \lesssim 1400$ GeV. The stop becomes lighter
than the lightest neutralino and it is excluded by our requirement
that allows only the solutions for which the lightest neutralino is
the LSP. While the LSP stop bounds the $\mu'-$term from below as
$\mu' \gtrsim -\mu$, this bound can be found different if some other
sparticles become LSP.

\begin{figure}[h!]
\includegraphics[scale=1,height=4.6cm,angle=0]{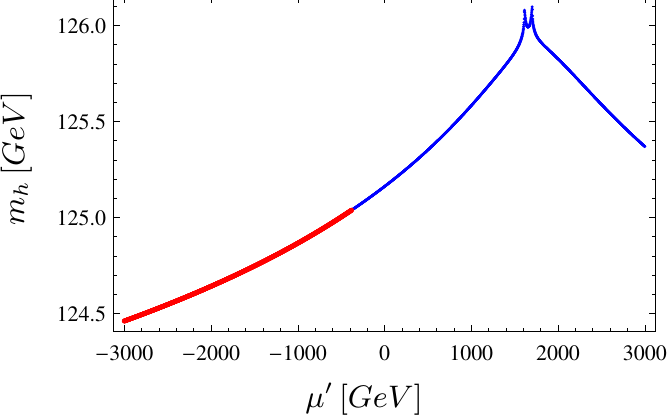}
\hfill
\includegraphics[scale=1,height=4.6cm,angle=0]{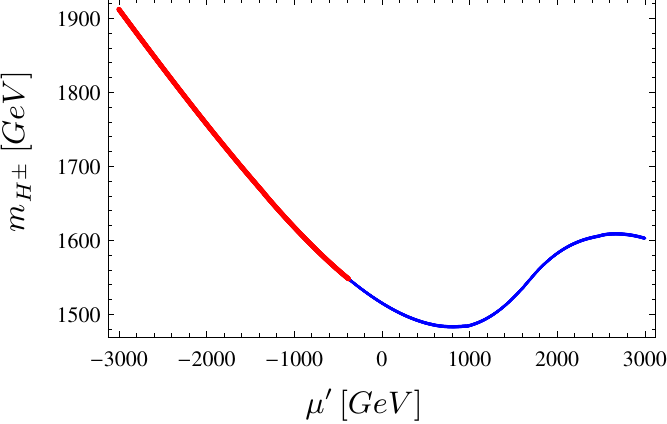}
\caption{Plots in $m_{h}-\mu'$  and $m_{H^{\pm}}-\mu'$ planes for
BMP1. The color coding is the same as Figure \ref{figure1}.} \label{figure4}
\end{figure}

Before concluding this section, sensitivity of the Higgs sector to
the $\mu'-$term should be investigated.As emphasized above the
$\mu'-$term dominantly controls the Higgsino masses, and
hence one can expect a different phenomenology associated
with the physical Higgs states of MSSM because of the loop level contributions from the Higgsinos. Figure \ref{figure4}
displays the results in  $m_{h}-\mu'$  and $m_{H^{\pm}}-\mu'$ planes
for BMP1. The
color coding is the same as Figure \ref{figure1}. In contrast to the
expectation, the SM-like Higgs boson mass decreases only $\sim 0.5$
GeV as $\mu'$ increases in its negative values in the red region. On
the other hand, the other Higgs states, which are rather heavy, seem
more sensitive to the $\mu'-$term. Related with heavy higgses,
$m_{A}$, $m_{H}$ and $m_{H^{\pm}}$ exhibit similar behavior, and
hence we present our results only in $m_{H^{\pm}}-\mu'$ plane.
According to the plot obtained, masses of these heavy Higgs states
increase with $\mu'$, and it is possible to rise their masses up
about 400 GeV in the red region for BMP1.

\begin{figure}[h!]\hspace{-2.0cm}
\begin{center}
\includegraphics[scale=0.7]{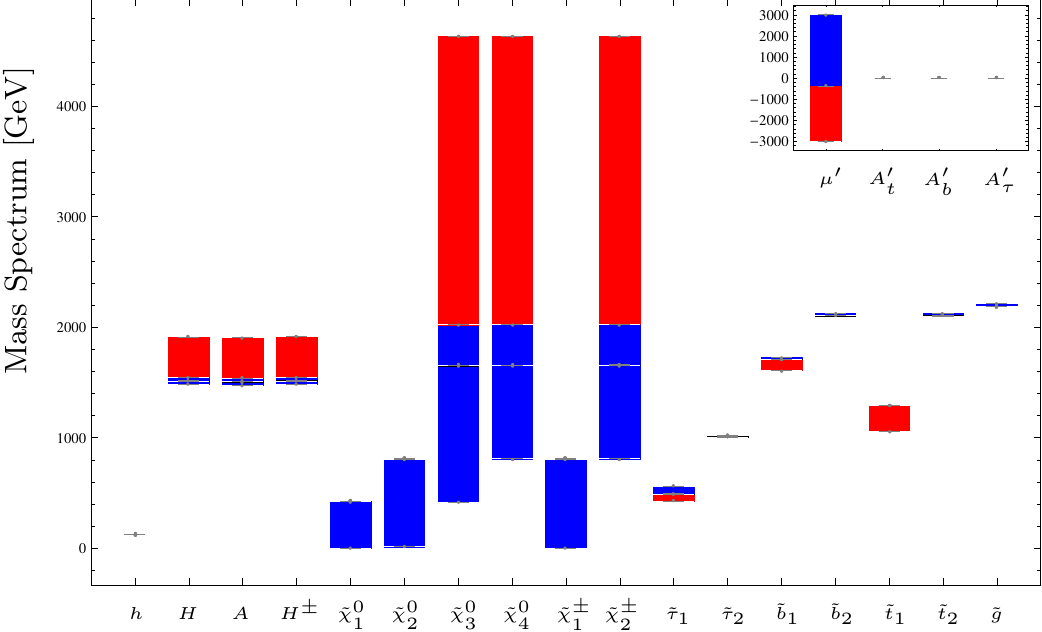}
\hfill
\includegraphics[scale=0.7]{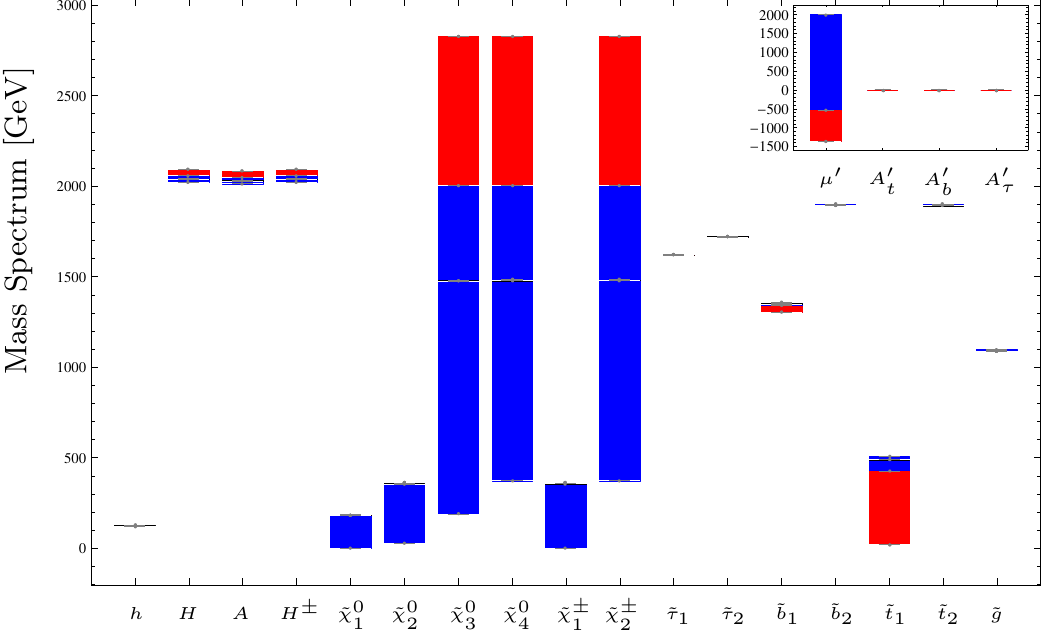}
\end{center}
\vspace{-0.4cm} \caption{Mass spectrum of the MSSM against
$\mu^\prime$ for BMP1 (left) and BMP2 (right) panels. Our color
coding is as in Figure \ref{figure1}.} \label{figure5}
\end{figure}

We have observed very similar behavior of the low scale observables under the presence of NH terms, and hence we do no repeat all the results for BMP2.

In Figure \ref{figure5}, in order to sum up our findings, we present
two charts which show the changes in supersymmetric mass spectra for
both BMP1 and BMP2. We use the same color coding as that we use in
the plots. While the bars show the total changes in masses, red
represents the masses consistent with the experimental constraints
including those from the rare decays of B-meson. The left chart
represents BMP1, while the right one displays BMP2. These two charts
clearly exemplify the similar behavior under the presence of the NH
terms. The small charts at the right top of the big ones represent
the scan over the NH parameters with the same color coding. A larger
range for $\mu'$ is found for BMP1 than BMP2, since the LSP stop is
excluded in the case of BMP2. As mentioned above, masses of the
heavy Higgs boson states change with the $\mu'-$term in the same
amount, while the change in the SM-like Higgs boson is negligible in
the charts. The neutralino and chargino sector represent the
interchange between the higgsinos and bino-wino. The red region in
the two lightest neutralinos and similarly in the lightest chargino
is not visible, since their masses are not changed by $\mu'$ in the
red region. A small change in the lightest sbottom is observed,
while it is at the order of a few hundred GeV in the lightest stop.
Besides this, masses of heavy sbottom and stop states negligibly changes. Finally gluino mass receives no contribution at all, as should be expected.

\subsection{$A'_{t,b,\tau}$ terms}
\label{Aprimes}
In the previous section, we have considered the NH contributions only from $\mu'$. It is because the most significant contributions to the B-physics observables come from $\mu'$.
Even though NH $A'_{t,b,\tau} $ terms are effective, their contributions are not enough to correct the results for the targeted decays of B-meson, at least
for the selected values of our parameters in BMP1. This is not a must and the situation might be different in alternative selections.

\begin{figure}[h!]
\begin{center}
\includegraphics[scale=1,height=3.1cm,angle=0]{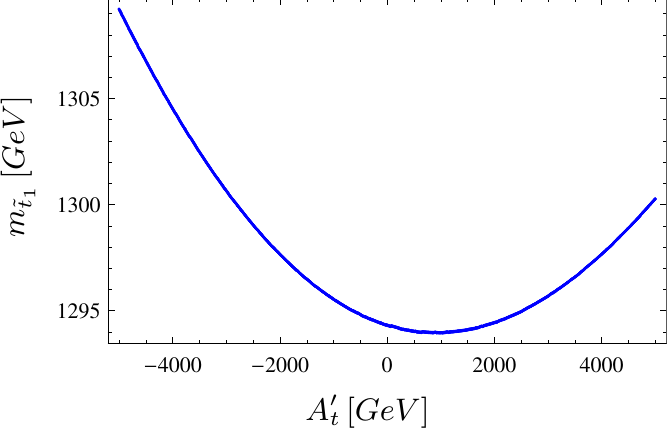}
\hfill
\includegraphics[scale=1,height=3.1cm,angle=0]{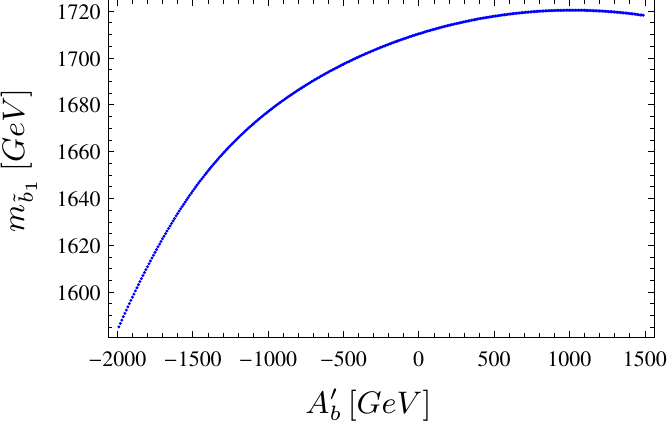}
\hfill
\includegraphics[scale=1,height=3.1cm,angle=0]{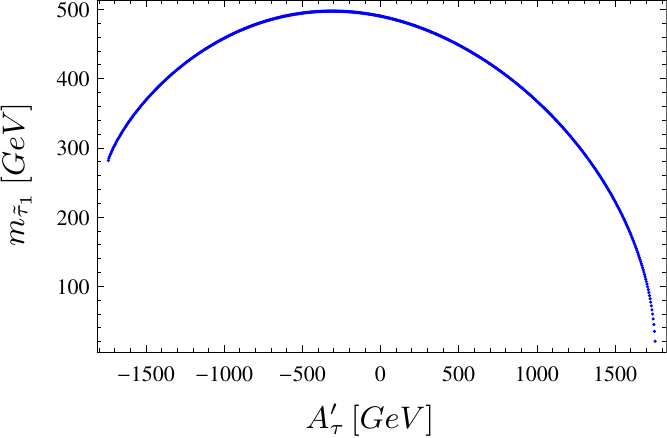}
\end{center}
\caption{Plots in $m_{\tilde{t}_{1}}-A'_{t}$, $m_{\tilde{b}_{1}}-A'_{b}$ and $m_{\tilde{\tau}_{1}}-A'_{\tau}$ planes. The curves are all in blue, since all results are excluded by some of the constraints. Each plot is obtained by varying only a single parameter that is represented on the x-axis of the planes.}
\label{figure6}
\end{figure}

 Let us start with Figure \ref{figure6} where we present our results in $m_{\tilde{t}_{1}}-A'_{t}$, $m_{\tilde{b}_{1}}-A'_{b}$ and $m_{\tilde{\tau}_{1}}-A'_{\tau}$ planes. The curves are all in blue, since all results are excluded by the constraints from the rare decays of B-meson and higgs mass measurements. Each plot is obtained by varying only a single parameter that is represented on the x-axis of the planes. The $m_{\tilde{t}_{1}}-A'_{t}$ plane shows that the effect of $A'_{t}$ is rather increasing the stop mass. The stop mass curve becomes steeper for negative values of $A'_{t}$. On the other hand, sbottom and stau masses exhibits opposite behavior under the NH effects. Sbottom mass is almost constant for the positive $A'_{b}$, and it decreases with increasing $A'_{b}$ in its negative values, while the stau mass decreases with both negative and positive values of $A'_{\tau}$. BMP1 predicts the LSP neutralino mass to be about 425 GeV, and as is seen from the $m_{\tilde{\tau}_{1}}-A'_{\tau}$ plane, stau becomes lighter than the LSP neutralino when $A'_{\tau} \gtrsim 700$ GeV that is excluded by our requirement that the lightest neutralino is always LSP.

Since one of the important and strict constraints comes from the observation of 125 GeV Higgs boson, one should also consider the NH trilinear
 impact on the Higgs mass. As is well-known, the SM-like Higgs boson mass is bounded by $M_{Z}$ from above, and one needs to use the two-loop level contributions in order to raise the Higgs boson mass
  up to 125 GeV. In the loop contributions, the third family of charged sfermions have a special importance, since their couplings to the Higgs boson are large in comparison to the first two families. `However, the mixing in sfermion sector behaves different depending on the flavor. In the case of staus and sbottoms it is proportional to  $ -(\mu\tan\beta + A'_{b,\tau}v_{u})$ and it is enhanced by the $\tan\beta$ parameter. Moreover, the negative sign in mixing of staus and sbottoms with $\tan\beta$ enhance can destabilize the Higgs potential, and this situation severely constrains the effects of $A'_{b,\tau}$ along with $\mu\tan\beta$ \cite{Carena:2012mw}. On the other hand, the mixing of stops is found as $\mu\cot\beta + A'_{t}v_{d}$. Note that $v_{d}$ behaves like $1/\tan\beta$. Despite its negative sign, the mixing in the stop sector exhibits $1/\tan\beta $ suppression, and hence it has more freedom to satisfy the vacuum stability constraint. Note that this discussion does not hold for the holomorphic $A_{t}$ term, since its effect is enhanced by $\tan\beta$, and it is constrained by the charge and color breaking minima as well \cite{Ellwanger:1999bv}.
  
\begin{figure}[h!]
\begin{center}
\includegraphics[scale=1,height=3.1cm,angle=0]{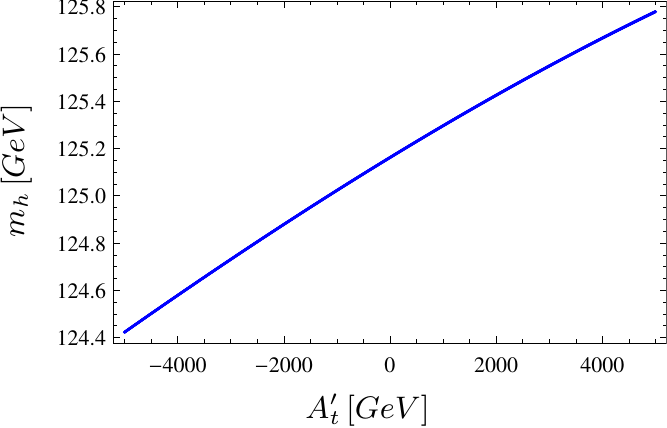}
\hfill
\includegraphics[scale=1,height=3.1cm,angle=0]{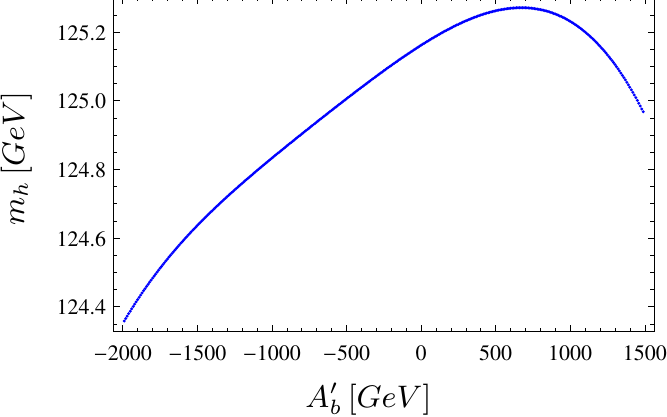}
\hfill
\includegraphics[scale=1,height=3.1cm,angle=0]{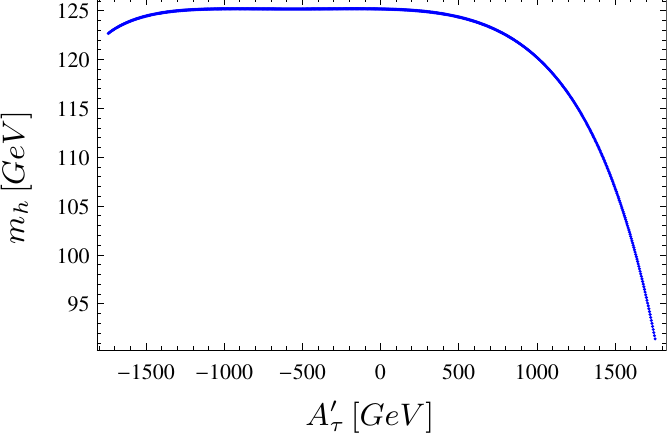}
\end{center}
\caption{Impact of the trilinears $A^\prime_t$, $A^\prime_b$ and $A^\prime_\tau$ on the lightest Higgs mass $m_h$ in BMP1.}
\label{figure7}
\end{figure}

 From Eqs.(\ref{defs}), the NH trilinear couplings, $A'_{t}, A'_{b}$ and $A'_{\tau}$ contribute respectively $-v_{d}A^{'\dagger}_{t}$ to the stop mixing, $v_{u}A^{'\dagger}_{b}$ to the sbottom mixing, and $v_{u}A^{'\dagger}_{\tau}$ to the stau mixing to be consistent with the 125 GeV Higgs boson mass. These contributions may relax the requirement of heavy sfermions or large mixings. Figure \ref{figure7} shows the impact of the trilinears $A^\prime_t$, $A^\prime_b$ and $A^\prime_\tau$ on the lightest Higgs mass $m_h$ in BMP1. The results for the NH trilinear contributions to the SM-like Higgs boson mass show that the significant contributions come from $A'_{t}$. The $m_{h}-A'_{t}$ plane shows a linear correlation between the SM-like Higgs boson mass and $A'_{t}$.
 In addition, $A'_{b}$ has nonzero contribution, but its contribution is minor  compared to $A'_{t}$. The $m_{h}-A'_{\tau}$ represents an interesting curve. The contribution from $A'_{\tau}$ is negligible for $-2000 \lesssim A'_{\tau} \lesssim 700$ GeV, and afterwards the mass curve makes a steep fall to $m_{h}\approx 90$ GeV. Recall that the stau becomes LSP in this region and hence it is excluded. Therefore, $A'_{\tau}$ has almost zero contribution to the SM-like Higgs boson mass in its allowed range.

\begin{figure}[h!]
\begin{center}
\includegraphics[scale=1,height=6cm,angle=0]{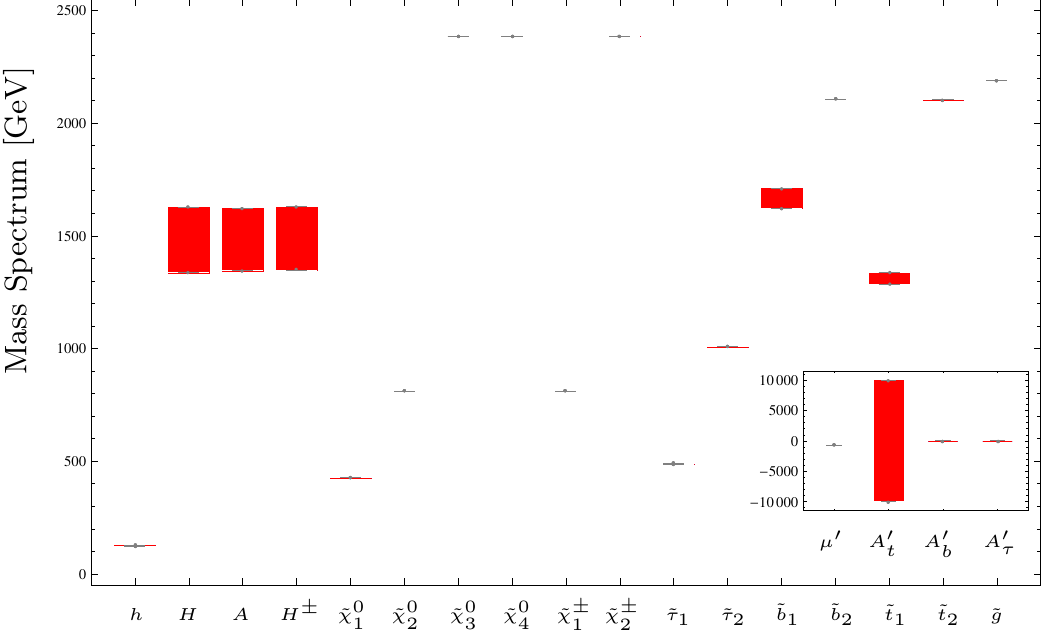}
\hfill
\includegraphics[scale=1,height=6cm,angle=0]{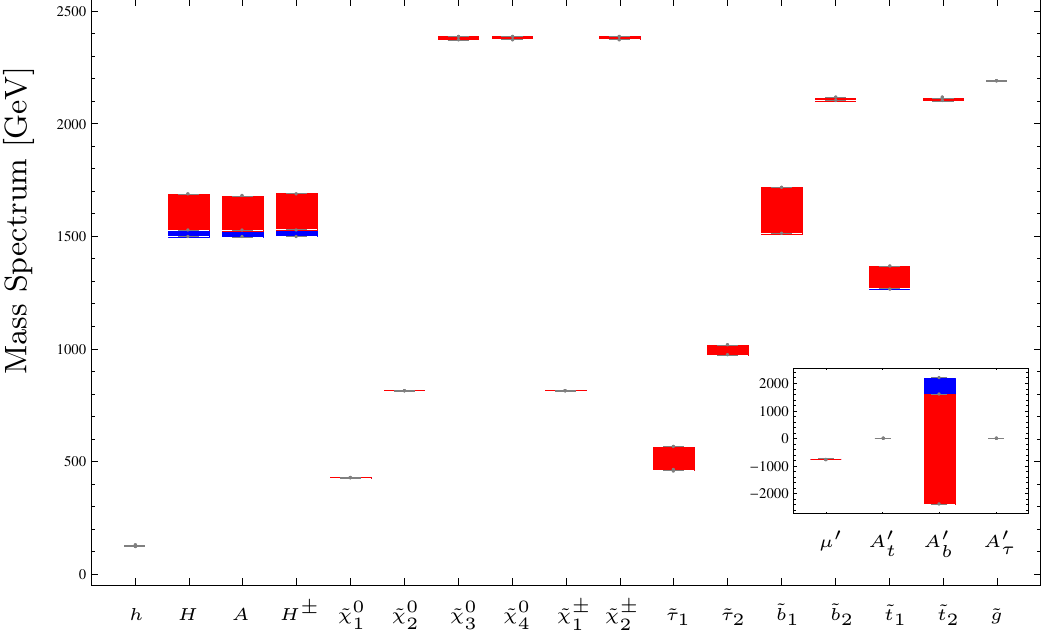}
\hfill
\includegraphics[scale=1,height=6cm,angle=0]{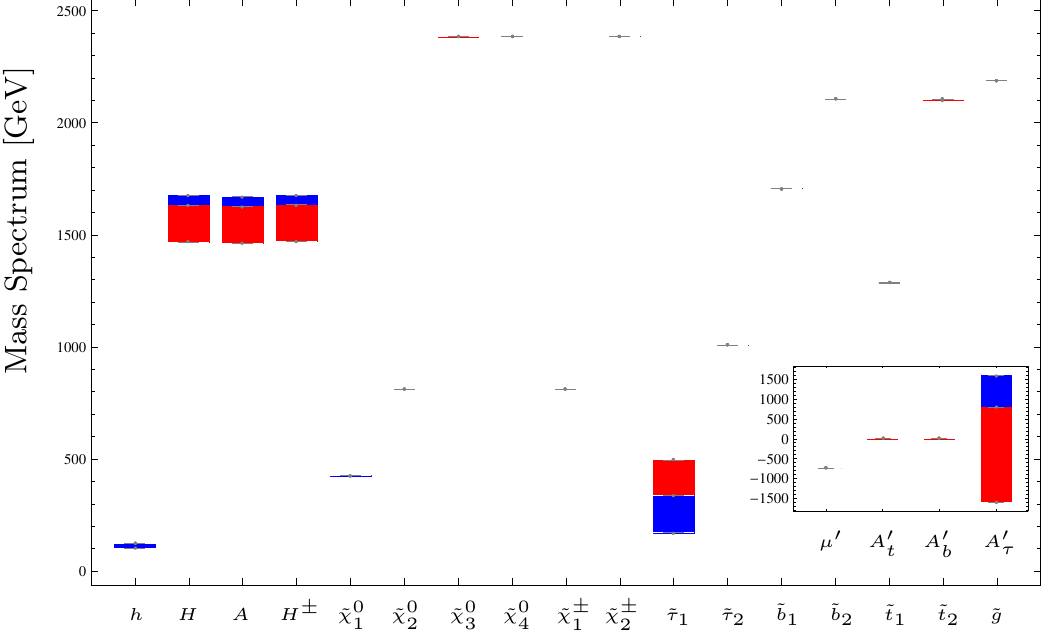}
\end{center}
\caption{Mass charts for $A'_{t}$, $A'_{b}$ and $A'_{\tau}$ with $\mu' = -750$ GeV respectively from top to bottom for BMP1. The color coding and explanation are the same as Figure \ref{figure5}.}
\label{figure8}
\end{figure}

In order to explicitly show the allowance and exclusion in ranges of the NH trilinear couplings, it is better to set $\mu'$ nonzero such that the results become consistent with all the experimental constraints mentioned in Section \ref{sec:scan}. For this purpose we choose a moderate value for $\mu'$ and set it to $-750$ GeV which contributes enough to satisfy all the constraints. We sum up our findings for the NH trilinear couplings in mass charts for $A'_{t}$, $A'_{b}$ and $A'_{\tau}$ with $\mu' = -750$ GeV respectively from top to bottom for BMP1 given in Figure \ref{figure8}. The color coding and explanation of the charts are same as in Figure \ref{figure5}.

The top chart represents the effects of $A'_{t}$, and it seems that once the constraints are satisfied, the contributions from $A'_{t}$ does not violate them despite its wide range. On the other hand, the contributions from $A'_{b}$ can contradict with the B-physics observables even if its range is not as wide as $A'_{t}$. In the case of $A'_{\tau}$, the blue part is excluded by the LSP neutralino requirement as mentioned above. It is a peculiar feature of our BMP1 that the stau and neutralino can be made nearly degenerate. If we  consider BMP2 instead of BMP1, LSP neutralino requirement would exclude some contributions from $A'_{t}$, since it leads to LSP stop at some point.

While contribution to the SM-like Higgs boson mass is not visible in the charts, the heavy Higgs boson states exhibit the same behavior as obtained in the chart given in Figure \ref{figure5} for $\mu'$. As expected, each NH trilinear coupling has a straightforward effect on the related particle. Namely, the impact of $A'_{t}$ on the stop mass, the impact of $A'_{b}$ on the sbottom mass, and the impact of $A'_{\tau}$ on the stau mass can be seen straightforwardly from the charts. However, they might behave differently. The stop tends to be heavier with the contributions from $A'_{t}$, while sbottom and stau become lighter in the case of nonzero $A'_{b}$ and $A'_{\tau}$ respectively. It is interesting to note that sbottom mass receives some contributions from $A'_{t}$ as well as from $A'_{b}$. It is because the threshold corrections to $Y_{b}$ partly depends on the stop mass at $M_{{\rm SUSY}}$ \cite{Pierce:1996zz}, where $M_{{\rm SUSY}}$ is the scale at which the supersymmetric particles decouple. Similarly, the stop mass can be changed with the contributions from $A'_{b}$ because of the threshold corrections to $Y_{t}$, but its change is not as much as that in the sbottom mass \cite{Pierce:1996zz}.

As can be predicted from the presented examples, besides the
stop-top degeneracy, one can predict novel sfermion decay patterns
which may be subject of future studies. It should be stressed for our NH terms that we assumed third family dominance $i.e.$ $A_u=A_t$, in this work, which is in fact a $3\times
3$ matrix with 9 entries in the CP conserving case. On the other hand, by
considering nonzero values for all families, one can study enhanced flavor
phenomenology, too.

\section{Note on Fine-Tuning}
\label{sec:finetuning}
The NH terms mingle the sparticles such that $H_{u}$ can couple to d-type quarks and charged leptons at tree level, while it also provides a vertex that $H_{d}$ couples to up-type quarks and we saw in previous sections that they could significantly change the phenomenology at the low energy scale. In addition to the experimental constraints, one could define also some phenomenological conditions such as LSP neutralino applied in our analysis. Besides  the experimental constraints and phenomenological conditions, one can also consider the fine-tuning in NHSSM, since it has more parameters which are involved in calculation of the low scale observables.

The measure of fine-tuning can be defined by considering the mass of Z-boson. Even though it is measured experimentally, it can be written in terms of the fundamental parameters obtained by minimizing the Higgs potential in NHSSM, as follows: 

\begin{equation}
\frac{1}{2}M_{Z}^{2} = -\mu^{2}+\frac{(m_{H_{d}}^{2}+\Sigma_{H_{d}})-(m_{H_{u}}^{2}+\Sigma_{H_{u}})\tan^{2}\beta}{\tan^{2}\beta - 1}
\label{Z-mass}
\end{equation}
where $\mu$ is the bilinear mixing term, $\tan\beta = \left\langle H_{u}\right\rangle / \left\langle H_{d}\right\rangle$, $m_{H_{u,d}}^{2}$ are the SSB mass terms of the Higgs doublets, $\Sigma_{H_{d}}$ and $\Sigma_{H_{u}}$ are the radiative corrections to the SSB mass terms of the Higgs doublets. Amount of the fine-tuning required to be consistent with the electroweak scale ($M_{{\rm EW}}\sim 100$ GeV) can be calculated by defining \cite{Baer:2012mv}

\begin{equation}
\Delta_{{\rm EW}} \equiv {\rm Max}(C_{i})/(M_{Z}^{2}/2)
\label{fine-tuning}
\end{equation}
where
\begin{equation}
C_{i}\equiv \left\lbrace
\begin{array}{c}
\hspace{-1.2cm}C_{H_{d}}=\mid m^{2}_{H_{d}}/(\tan^{2}\beta -1) \mid \\ \\
C_{H_{u}}=\mid m^{2}_{H_{u}}\tan^{2}\beta/(\tan^{2}\beta -1) \mid \\ \\
\hspace{-3.5cm}C_{\mu}=\mid -\mu ^{2}\mid
\end{array}
\right.
\label{fineterms}
\end{equation}

Comparing with the holomorphic MSSM framework, the minimization of the Higgs potential in NHSSM yields the same relation between the model parameters and $M_{Z}$. This follows from the fact that the higgsinos do not directly interfere in the scalar Higgs potential, and hence one can derive the same expressions for the fine-tuning measures as given in Eq.(\ref{fineterms}). On the other hand, as is mentioned above, the NH terms contribute to the observables also at loop levels, and hence the loop contributions $\Sigma_{H_{d}}$ and $\Sigma_{H_{u}}$ would be different in the NHSSM framework. The calculation of low scale parameters already include the loop contributions, and hence $C_{H_{d}}$ and $C_{H_{u}}$ in Eq.(\ref{fineterms}) are defined only with the SSB mass terms of the Higgs fields. Since these factors are suppressed by $\tan\beta$, the fine-tuning is mostly measured by the term $C_{\mu}$, and the NH terms do not have significant effects in the fine-tuning.

Even though the NH terms do not change the fine-tuning measurements, they can change the phenomenology in the regions which yield acceptable fine-tuning ($\Delta_{EW}\lesssim 1000$), when it is considered with mass spectrum of the supersymmetric particles. The benchmark points given in Table \ref{MSSMbenchmarks} are both acceptable under the fine-tuning requirement, since $\Delta_{EW}=661.5$ for BMP1, and $\Delta_{EW}=525.4$ for BMP2. The fine-tuning requirement bounds the stop mass as $m_{\tilde{t}_{1}} \gtrsim 300$ GeV \cite{Hall:2011aa}, and as seen from the Table \ref{MSSMbenchmarks}, the stop mass is about 500 GeV when the NH terms are absent. On the other hand, as is shown in Figure \ref{figure3},  the stop mass for BMP2 can be found as low as 180 GeV for $\mu'\sim - 1200$. At this point the stop is nearly degenerate with the top quark and distinguishing  $\tilde{t}_{1}\tilde{t}^{*}_{1}$ events at LHC is challenging, since such events can result in the identical final states with $t\bar{t}$, and the cross section of $\tilde{t}_{1}\tilde{t}^{*}_{1}$ is found to be less than the error in calculation of top pair production whose measure is given in Eq.(\ref{toppair}). Such light stops can hide in the top quark backgrounds in colliders, and they can escape from the observation. A recent study has shown that a very narrow region with $m_{0}\sim 9$ TeV, $M_{1/2}\sim 0.3$ TeV, $A_{0}\sim -18$ TeV and $\tan\beta \sim 34$ in the CMSSM parameter space can yield the light stop of mass $\lesssim 300$ GeV, and this region is highly fine-tuned ($\Delta_{EW}\sim 10000$) \cite{Demir:2014jqa}. Comparing our results displayed in Figure \ref{figure3} with those revealed in Ref. \cite{Demir:2014jqa}, the light stop region in the CMSSM can be enlarged. Furthermore, since the NH terms do not change the required amount of fine tuning, $\Delta_{EW}$ remains about 500, and hence, the light stop region in NHSSM can be realized with a reasonable amount of fine-tuning.

\section{Conclusion}
\label{sec:conc}

In this work, we studied the mass spectrum of MSSM with
new NH soft breaking terms. In doing this we respected the
experimental constraints especially from the rare decays of B-meson
and the mass bounds on the supersymmetric particles. We have chosen
two benchmark points from the CMSSM parameter space that are
currently excluded by the experimental results on the B-physics
observeables. By probing the impact of the NH
terms based on these two benchmark points we have deduced that the
enlarged soft supersymmetry breaking sector with the NH
terms has many advantages. 

First of all, the B-physics predictions of
CMSSM can be corrected with the contributions from the
NH terms and hence the CMSSM parameter space allowed by
the current experimental constraints can be found significantly
larger, if one performs a more detailed scan over its fundamental
parameters. Their contributions also change the mass spectrum of the
supersymmetric particles. We have find that the Higgs sector except
the SM-like Higgs boson exhibits a large sensitivity to the NH
terms. While the effects on the SM-like Higgs boson is negligible,
the masses of heavy Higgs states can differ up to 400 GeV. Among the
NH terms, $\mu'$ strongly controls the Higgsino masses and it leads
to Higgsino-like neutralino LSP whose mass is almost zero when $\mu'
\approx \mu$. This region also results in almost massless chargino
which is excluded by the LEP mass bound on the chargino. Besides the
Higgs sector, also the lightest stop, sbottom and stau are sensitive
to the NH contributions, while the heaviest states of them are
totally blind to the NH terms. Changes in the mass spectra can yield
different NLSP species such as stop and stau as we obtained for BMP1
and BMP2 and each NLSP has its own phenomenology.

In addition to NH enrichment in the low scale
phenomenology, we observe that the SM-like Higgs boson of mass about 125 GeV can be realized even when the stop mass is not too heavy, $m_{\tilde{t}_{1}}\sim 180$ GeV, in contrast to CMSSM without NH terms \cite{Czakon:2014fka}. Lowering the stop mass brings up the discussion about the light stop mass nearly degenerate to the top quark. In this case the stop can hide in the top quark background and escape from the observation in colliders. In the CMSSM framework with NH terms, we realize that such light stops can be consistent with the experimental constraints, and light stop regions can remain reasonably fine-tuned ($\Delta_{EW}\lesssim 1000$). In this context, the NH soft breaking terms can provide a reasonable resolution to the naturalness problem by lowering the sparticle masses.

The allowed ranges for some of the NH terms are striking since they can be as large as hundreds of GeV and satisfy all the criteria we have considered. The excluded regions can receive significant contributions such that the most constrained supersymmetric models such as CMSSM still offer testable solutions. Under the pressure from the current experimental results, it seems
crucial to consider MSSM and its alternative extensions, and the
results presented in our study is an existential example of
additional NH soft breaking terms possibility, which could be
improved with a more through analysis.

\section{Acknowledgments}
We would like to thank Florian Staub first for preparing NHSSM package. We also express our gratitude to Dimitri I. Kazakov and Durmu\c{s} A. Demir for their useful discussions and comments. This work is partly supported by Russian
Foundation for Basic Research (RFBR) grant No. 14-02-00494-a (\c{S}HT).

\end{document}